\documentclass{ws-ijmpa}

\begin{document}

\markboth{Dawson, Jackson, Reina and Wackeroth}
{Higgs Boson Production with Bottom Quarks at Hadron Colliders}

%%%%%%%%%%%%%%%%%%%%% Publisher's Area please ignore %%%%%%%%%%%%%%%
%
\catchline{}{}{}{}{}
%
%%%%%%%%%%%%%%%%%%%%%%%%%%%%%%%%%%%%%%%%%%%%%%%%%%%%%%%%%%%%%%%%%%%%

\title{Higgs Boson Production with Bottom Quarks\\
at Hadron Colliders}

\author{\footnotesize S. DAWSON}

\address{Physics Department, Brookhaven National Laboratory,\\
Upton, NY 11973-5000,
USA}

\author{\footnotesize C.B. JACKSON\footnote{Talk given 
by C.B. Jackson} \mbox{\,\,\,\,}and L. REINA}

\address{Physics Department, Florida State University,\\
Tallahassee, FL 32306-4350, USA
}
%
%\author{\footnotesize L. REINA}
%
%\address{Physics Department, Florida State University,\\
%Tallahassee, FL 32306-4350, USA
%}

\author{\footnotesize D. WACKEROTH}

\address{Department of Physics, SUNY at Buffalo,\\
Buffalo, NY 14260-1500, USA
}

\maketitle

%\pub{Received (Day Month Year)}{Revised (Day Month Year)}

\begin{abstract}
%The production of a Higgs boson with a pair of $b\bar{b}$ quarks can play
%a crucial role for discovery at both the Tevatron and the Large Hadron
%Collider (LHC).  In some extensions of the Standard Model (SM), for example 
%the Minimal Supersymmetric Standard Model (MSSM), the Yukawa coupling of the 
%$b$ quark becomes strongly enhanced causing the production of Higgs bosons
%with $b$ quarks to be a dominant production process.  In this talk, we present
%results for the cross section calculations including next-to-leading order 
%(NLO) QCD corrections for $p\bar{p},pp \to b\bar{b}h$.  We also the review 
%the status of the comparison 
%between the two theoretical schemes in which the production of Higgs bosons 
%with bottom quarks can be calculated.    
%
We present results for the production cross section of a Higgs boson with
a pair of $b\bar{b}$ quarks, including next-to-leading order (NLO) QCD 
corrections.

\keywords{Higgs; MSSM.}
\end{abstract}

\section{Introduction}	%) A SECTION HEADING

One of the most pressing issues at current and future hadron collider
experiments is the discovery of one (or more) Higgs bosons.  The 
Standard Model (SM) predicts a single Higgs boson whose couplings to fermions
are proportional to the mass of the fermion, hence the production of the
Higgs boson with light quarks (e.g. bottom quarks) is suppressed compared 
to other production channels involving top quarks or heavy gauge bosons.  
However, in some extensions of the SM (such as the Minimal Supersymmetric
Standard Model, or MSSM), the 
Yukawa couplings for $b$ quarks can become strongly enhanced.  In this 
scenario, the production of a Higgs boson with $b\bar{b}$ can dominate
over other production channels and could provide a 
unique opportunity to directly probe the $b$ quark Yukawa coupling.

The theoretical prediction of $b\bar{b}h$ production at hadron colliders 
involves several subtle issues and depends on the number of $b$ quarks 
identified, or $\emph{tagged}$, in the final state.  Potentially large 
logarithms arise from the integration over the phase space of the final 
state $b$ quarks.  These large logarithms appear at all orders in 
perturbation theory and could spoil the convergence of the perturbative
expansion of total and differential cross sections.  Currently, there 
are two approaches to calculating the theoretical prediction of $b\bar{b}
h$ production.  Working under certain kinematic approximations, and 
adopting the so-called $\emph{five-flavor-number scheme}$ (5FNS), the 
logarithms 
can be resummed by using a bottom quark Parton Distribution Function
\cite{Barnett} $^{,}$ \cite{Olness}.  Alternatively,
working with no kinematic approximations, and adopting the so-called
$\emph{four-flavor-number scheme}$ (4FNS), one can compute the cross 
section for
$p\bar{p},pp \to b\bar{b}h$ at fixed order in
QCD without resumming higher order collinear logaritms \cite{Dittmaier}
$^{,}$
\cite{LH}.  

In this talk, we present the results of the total cross section calculation 
for $p\bar{p},pp \to b\bar{b}h$ at next-to-leading order (NLO) in 
QCD for two different scenarios: two high-$p_T$ $b$ jets and 
one high-$p_T$ $b$ jet 
corresponding to two or one $b$ jet(s) identified in the final
state. Results for the case of no high-$p_T$ $b$ jets have been reviewed
in a previous study \cite{LH}.   
For the case of one high-$p_T$ $b$ jet in the final
state, we compare our fixed-order result (4FNS) with the prediction 
obtained by using $gb \to bh$ (5FNS), when NLO QCD corrections have been
included in both cases \cite{bh}.

\section{Results for $b\bar{b}h$ Production}

In order to select the events with high-$p_T$ $b$ jets, we place kinematic cuts
on the transverse momentum, $p_T^{b,\bar{b}} >$ 20 GeV, and pseudorapidity,
$|\eta_{b,\bar{b}}| <$ 2 (2.5) at the Tevatron (LHC), of the final state 
$b$ quarks.  Results from the case of two high-$p_T$ $b$ jets are shown 
in Fig. 1 
\cite{exclusive}.  The
two sets of curves correspond to different renormalization schemes for the
bottom quark mass in the bottom quark Yukawa coupling.  In
both cases, the dependence on the renormalization and factorization schemes 
is greatly reduced by the inclusion of NLO QCD corrections.  It is 
interesting to note that the $\overline{MS}$ scheme exhibits better behavior 
at small scales and also seems to improve the convergence of the perturbative
series.
\begin{figure}
\centerline{\psfig{file=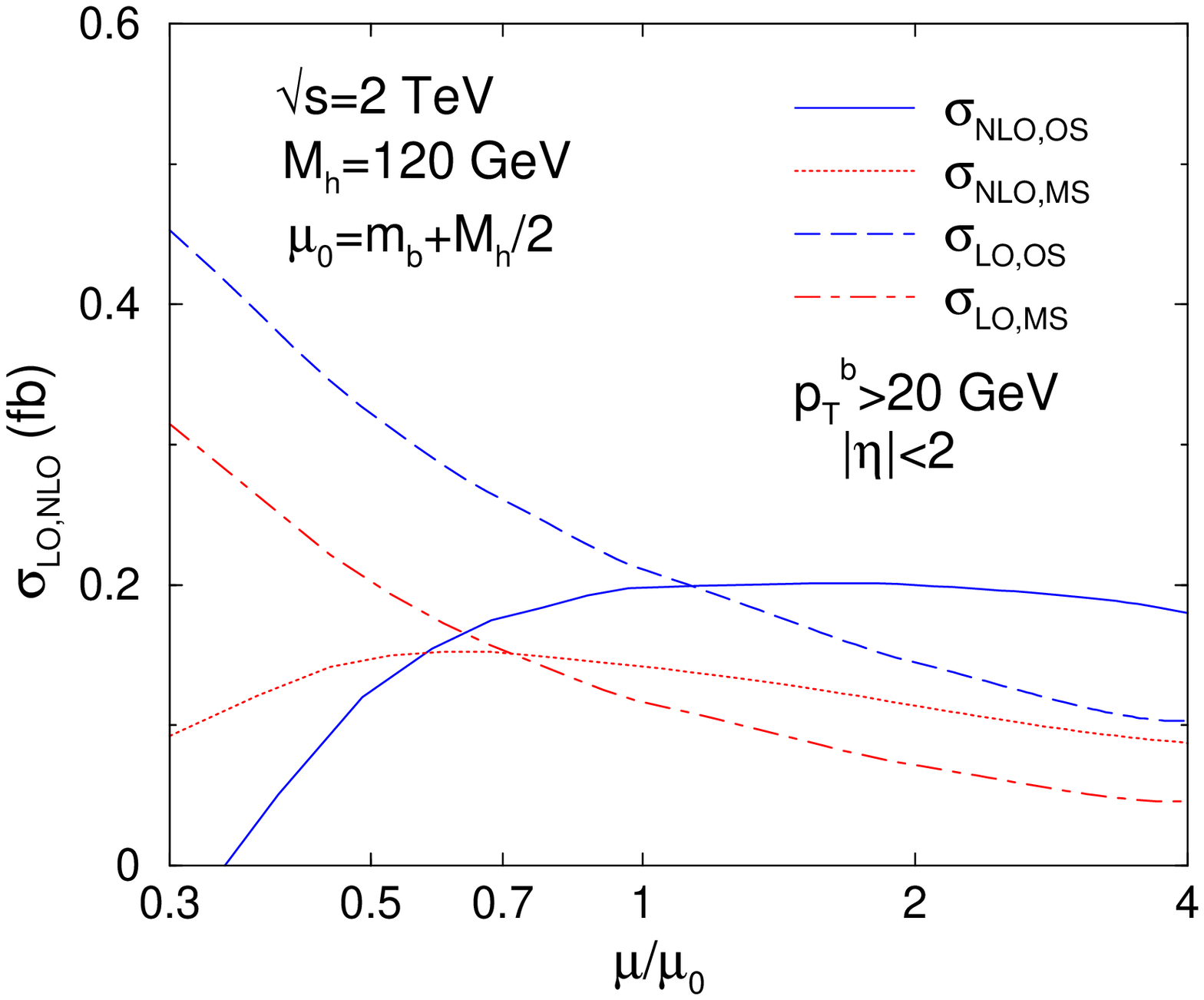,width=5cm},
\hspace{0.5cm} \psfig{file=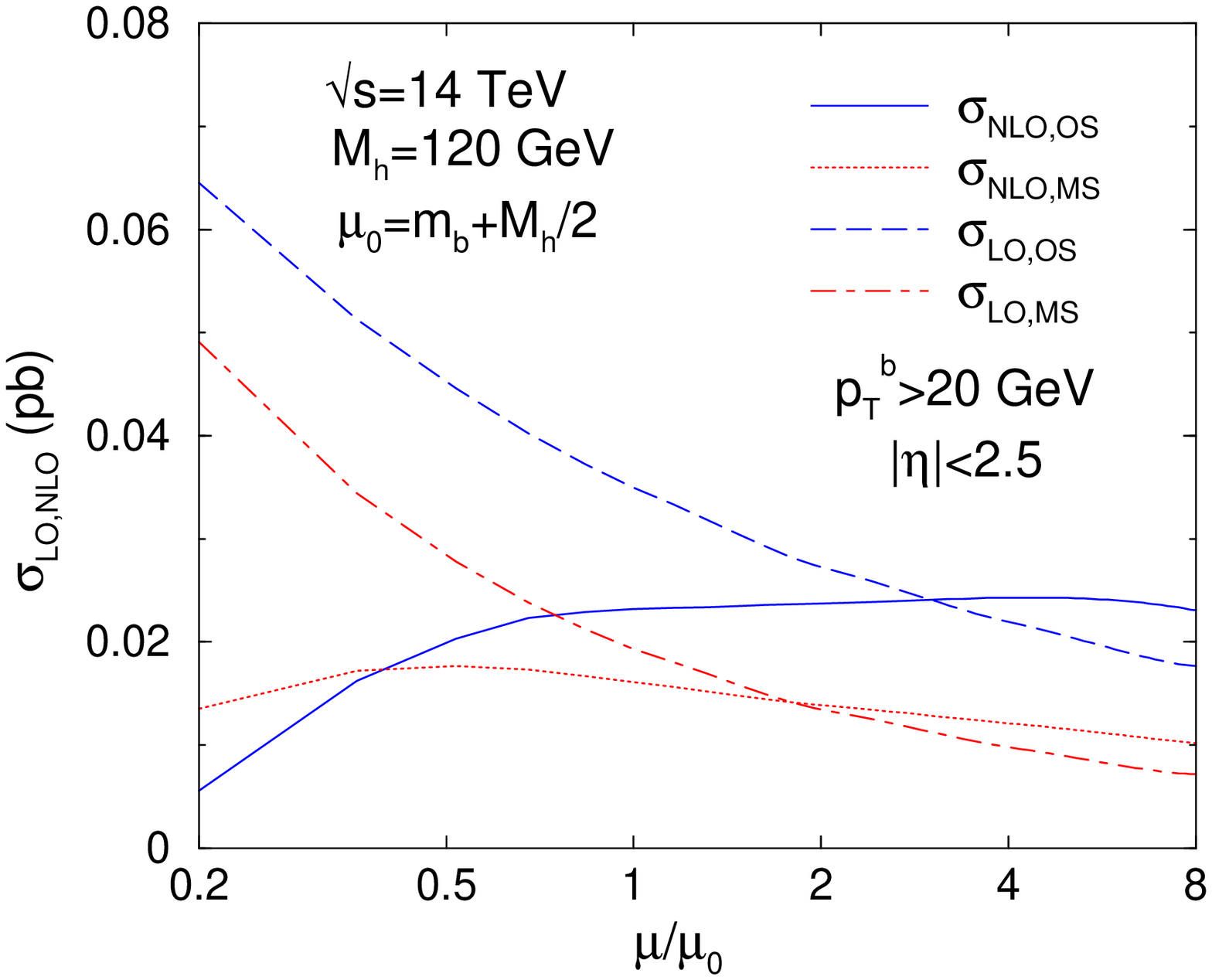,width=5.1cm}}
\caption{$\sigma_{NLO}$ and $\sigma_{LO}$ for $p\bar{p},pp \to b\bar{b}h$
with two high-$p_T$ $b$ jets at $\sqrt{s}$ = 2 TeV (left) and 
$\sqrt{s}$ = 14 TeV (right) as a function of the renormalization/factorization
 scale $\mu$, for $M_h$ = 120 GeV.}
\end{figure}
The calculation of the NLO cross section for $b\bar{b}h$ for 
the case of 
(at least) one high-$p_T$ $b$ jet has been performed in both the
4FNS \cite{Dittmaier}$^,$ \cite{LH} and the 5FNS \cite{Scott}(where the 
leading order 
process becomes $gb(\bar{b})
\to b(\bar{b})h$).  In a previous study \cite{LH}, the two cross
section calculations were found to be in agreement (within error), but the
5FNS calculation seemed to predict a slightly larger value for most of the
Higgs mass range.  In a subsequent paper \cite{bh}, we have investigated 
the inclusion of
diagrams containing top quark loops, that were previously 
neglected in the 5FNS, by implementing them into MCFM\cite{MCFM}.  
%The contribution from both diagrams is proportional to powers of the bottom
%mass and hence was neglected in the massless calculation of $gb(\bar{b})
%\to b(\bar{b})h$ 
%\cite{Scott}.  In any case, the bottom quark loop diagrams are essentially 
%suppressed by factors of ${\cal{O}}$$(\frac{m_b^2}{M_h^2})$ and, thus, are 
%truly neglible. 
The contribution from the top loop diagram(s) to the total cross section 
in the SM calculation can become
numerically important.  In fact, inclusion of these diagrams was found to 
reduce the 5FNS cross section 
by 15\%(10\%) at the Tevatron(LHC) over most of the Higgs mass range 
\cite{bh}.  
Results are shown in Fig.2 
for the comparison between the 4FNS and 5FNS schemes after the inclusion
of the top loop diagram in the latter calculation.  Clearly, within the 
theoretical 
uncertainties, the SM calculations in the 4FNS and 5FNS are in very good 
agreement.  The agreement is preserved in the MSSM with large $\tan\beta$,
since the top loop diagrams become negligible. 

\begin{figure}
\centerline{\psfig{file=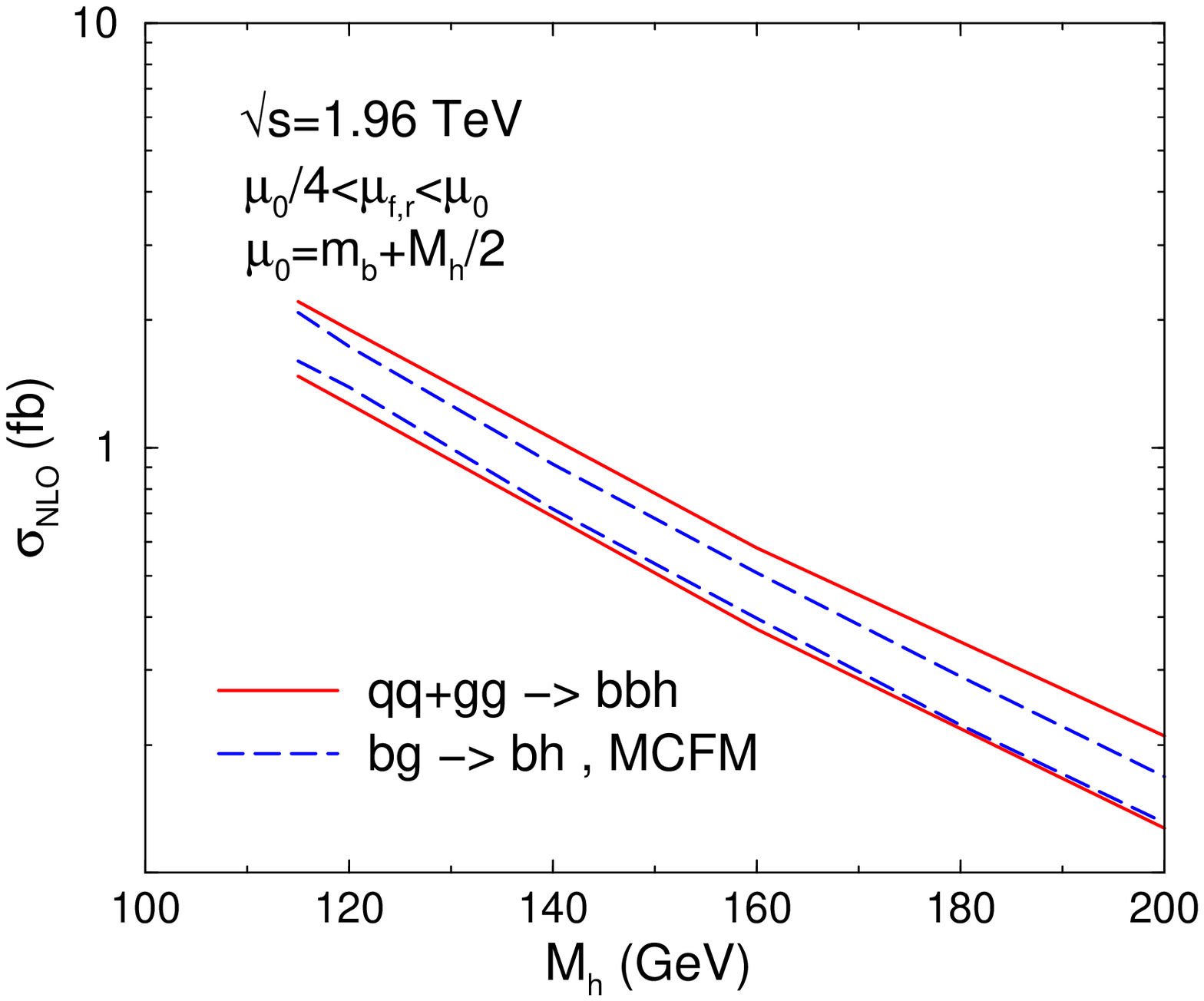,width=5cm},
\hspace{0.5cm} \psfig{file=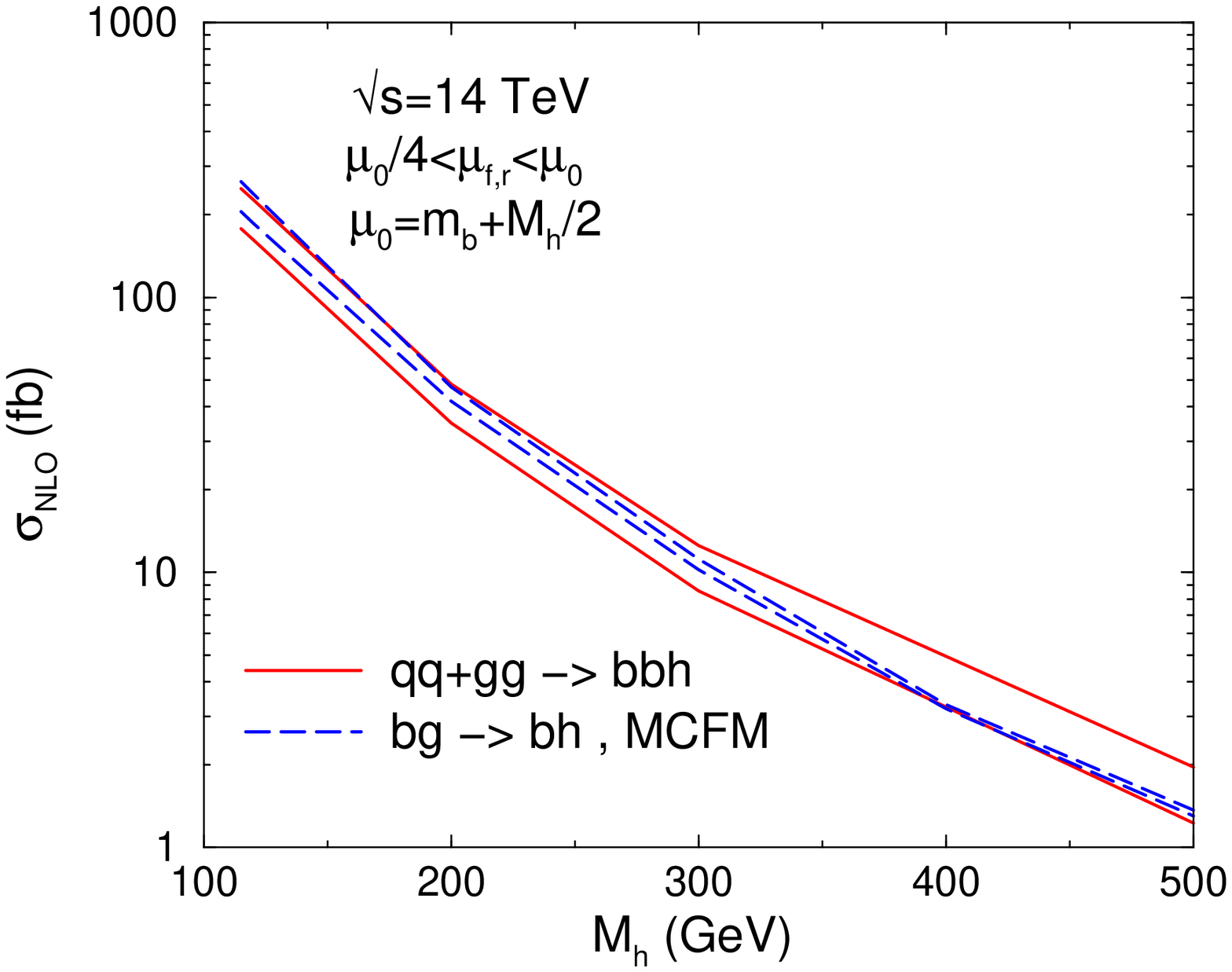,width=5.2cm}}
\caption{Total NLO cross section for $p\bar{p},pp \to b\bar{b}h$ with at
least one high-$p_T$ $b$ jet at the Tevatron(left) and LHC(right).  The
uncertainty bands are obtained by varying the renormalization and factorization
scales independently around the central value ($\mu_r$ = $\mu_f$ = $\mu_0/2$).
The solid(dashed) curves correspond to the 4FNS(5FNS) calculations.}
\end{figure}

\section*{Acknowledgments}

The work of S.D. (L.R., C.B.J.) and D.W. is supported in part by the U.S. 
Department of Energy under grant DE-AC02-76CH00016 (DE-FG02-97ER41022) and
by the National Science Foundation under grant NSF-PHY-0244875, respectively.

%\section*{References}

\end{document}